# Micro-tip manipulated origami for robust twisted few-layer graphene

Ruo-Jue Zou[†], Long Deng[†], Si-Min Xue, Feng-Fei Cai, Ling-Hui Tong, Yang Zhang, Yuan Tian, Li Zhang, Lijie Zhang, Zhihui Qin, and Long-Jing Yin*

*Key Laboratory for Micro/Nano Optoelectronic Devices of Ministry of Education & Hunan Provincial Key Laboratory of Low-Dimensional Structural Physics and Devices, School of Physics and Electronics, Hunan University, Changsha 410082, China*

[†]These authors contributed equally to this work.
*Corresponding author: yinlj@hnu.edu.cn

**Twisted few-layer graphene (tFLG) has emerged as an ideal model system for investigating novel strongly correlated and topological phenomena. However, the experimental construction of tFLG with high structural stability is still challenging. Here, we introduce a highly accessible method for fabricating robust tFLG by polymer micro-tip manipulated origami. Through using a self-prepared polymer micro-tip—which is composed of multiple dimethylpolysiloxane, poly(vinyl chloride), and graphite sheets—to fold graphene layers, we fabricated tFLG with different twist angles (0°-30°) and various layers, including twisted bilayers (1+1), twisted double-bilayers (2+2), twisted double-trilayers (3+3), and thicker layers. Even *ABC*-stacked tFLG were created, such as twisted *ABC/ABC* and *ABC/ABA* graphene coexisting in an *ABC-ABA* domain wall region. We found that the origami-fabricated tFLG exhibits high stability against thermal and mechanical perturbations including heating and transferring, which could be attributed to its special folding and tearing structures. Moreover, based on the rich types of samples, we revealed twist-angle and stacking-order dependent Raman characteristics of tFLG, which is valuable for understanding the stacking-modulated phonon spectroscopy. Our experiments provide a simple and efficient approach to construct structurally robust tFLG, paving the way for the study of highly stable twisted van der Waals heterostructures.**

Twisted few-layer graphene (tFLG) has recently attracted extensive attention due to the emergence of flat and topological band structures near the Fermi level. A variety of strongly correlated and topological electronic phases, such as correlated insulator states, superconductivity, and anomalous quantum Hall effects, have been experimentally discovered in tFLG.[1-7] It was demonstrated that such quantum states are highly dependent on the twist angle and the stacking sequence between the component layers, making the system both highly tunable and sensitive.[8-10] At present, the major method of fabricating tFLG samples is to use the 'tear-and-stack' transfer technique.[11-14] During this technique, the twisting configuration is constructed by stacking isolated graphene sheets through mechanical (stacking and encapsulation) and thermal (annealing) processes.[15,16] These external perturbations, together with the technological operations in further device fabrications, can easily alter the value and uniformity of the intended twist angle due to the metastability of the twisting structure, resulting in instability of the prepared tFLG samples.[17,18] How to fabricate twisted graphene with high stability therefore remains highly desirable.

Here, we propose a strategy for constructing robust tFLG by forming a folding-twisted structure in a single, whole graphene flake using polymer micro-tip manipulated origami. Previously, introducing twisting in bilayer or tetralayer graphene has been achieved by STM or AFM tip folding,[19-25] but it relies on expensive equipment and lacks accessibility in device applications. Other methods, such as water flushing and ultrasound treatment,[26,27] have also been reported to fold single-layer graphene, but mostly require wet conditions or complex operations. We develop a folding-graphene technique under atmospheric environment by using a simple, self-prepared polymer micro-tip. Using this technique, we fabricated different types of tFLG, such as twisted bilayers (1+1), twisted double-bilayers (2+2) to double-ten-layers (10+10), with the twist angle ranging from 0° to 30°, and even twisted ABC-stacked graphene. The generated folded-tFLG are all connected by a folding boundary between the twisted layers and mostly contain tearing edges. We demonstrate that this specific folded-tFLG exhibits high structural robustness against thermal and mechanical disturbances, such as annealing (up to 500 °C) and transferring. In addition, based on the obtained diverse tFLG, their twist-angle and stacking-order dependent Raman characteristics are also revealed and discussed.

The polymer micro-tip was prepared by dimethylpolysiloxane (PDMS), poly(vinyl chloride) (PVC), and graphite flakes, with its configuration illustrated in Fig. 1(a). First, we continuously stacked solidified PDMS sheets with decreasing sizes on a glass slide to generate a micro-dome with a curvature of approximately several tens of micrometers. Subsequently, the micro-dome was covered by a commercial PVC layer (see Supplementary Material and Fig. S1 for details). This type of polymer micro-dome has been previously used to achieve 3D manipulation of 2D materials.[28] However, such manipulation usually handles thick materials with a thickness of ~25 nm-40 nm, which fails to meet our requirements for preparing tFLG. Here, we further optimize this polymer-based manipulation method for thinner 2D materials. Due to the PVC polymer exhibiting firm adhesion with 2D layered materials at finite temperatures,[28,29] we therefore used the PVC-covered micro-dome to pick up a small graphite flake, tens of micrometers size, at 70 °C and adhere it to the top, forming our final polymer micro-tip [Fig. 1(a) and see Fig. S2 in the Supplementary Material for more details]. Adhering a small-size graphite flake at the top of the polymer micro-tip can result in a smaller contact point with the target graphene layers, enabling more precise folding of few-layer graphene. In addition, the exposed surface of the graphite flake at the top of the prepared microneedle is the surface that contacts $SiO_2$ after mechanical exfoliation. This surface usually exhibits a very clean characteristic, which could make the manipulation of few-layer graphene more controllable and cleaner.

With the help of the developed polymer micro-tip, we can achieve efficient folding and twisting manipulations of thin graphene layers. Figure 1(b) shows the schematic of micro-tip-assisted origami operation of graphene. Briefly, we first contact the micro-tip to the position of $SiO_2$ surface that is close to the edge of a graphene sheet, and adjust the micro-tip height to regulate the contact area under an optical microscope. After adjusting to an appropriate contact area (several to tens of micrometers), we move the tip across the graphene sheet to fold it. Through this polymer micro-tip-manipulated method, we can fold graphene sheets with different layers. Figure 1(c)-(h) shows the optical images of graphene sheets with different layers before and after folding. A flat, folded few-layer graphene region can be clearly visualized (also see Fig. 2(a) for the thickness in the AFM height-profile[30]), with the folding region covering an area of about several tens of square micrometers. The thinnest graphene sheet that can be folded is single-layer graphene as shown in Fig. 1(c), and thicker few-layer graphene can also be fabricated as shown in Fig. 1(h). Besides, we can fold few-layer graphene with different stacking orders, such as ABC and ABA stacking (see below).

Based on the above micro-tip-manipulated folding technique, different layers tFLG with various twist angles can be achieved. The folding naturally results in the lattice misalignment between the two folded graphene parts and thus generates a twist. The twist angle $\theta$ can be determined by measuring the angle $\varphi$ between the folding boundary and the graphene straight edge, using the relation $\theta = 180°-2\varphi$,[23,31] as illustrated in Fig. 2(a) (see Fig. S3 in the Supplementary Material for details about the extraction of $\varphi$). Under unintentional folding, the generated twist angle exhibits a random distribution with its value ranging from 0° to 30°. Figure 2(b) shows the statistical histogram of the angles for the prepared 42 twisted double-bilayer graphene [t(2+2)LG] samples. It roughly follows a Gaussian distribution, with the angles of highest probability occurring around 10°-20°. It is worth pointing out that a specific twist angle, such as a small angle, could be achieved by folding graphene edge along a selective direction, allowing for the realization of custom-designed twisted graphene.

We now study the Raman characteristics of the folding-created tFLG. It has been reported that the Raman characteristics of graphene layers are highly dependent on the interlayer stacking orders.[32-34] Based on the obtained rich types of samples, we thus can conduct a systematic investigation for the twist-angle and stacking-order dependent Raman properties of tFLG. Figure 2(c) shows Raman spectra of a series of twisted (2+2) samples with various twist angles. The typical G and 2D modes are most evident. It is known that the 2D peak is the most sensitive peak to the changes in the electronic and phonon band structures of graphene. For example, its peak width, position, and intensity exhibit a complex dependence on the twist angle.[35] Figure 2(d) shows the relationship between the full width at half-maximum (FWHM) of the 2D peak and the twist angle for all of the folded-t(2+2)LG samples. The 2D-peak width exhibits a non-monotonic dependence on the twist angle, with its value being usually larger than that of normal bilayer graphene, especially for angles < 15° [also visualized in the 2D-peak-width mapping; see Fig. 2(e) for the 11° t(2+2)LG as an example]. There is a blueshift of the 2D-peak position [Fig. 2(f)], which shows a similar non-monotonic dependence on the twist angle as the peak width. In addition, the 2D-peak intensity [Fig. 2(g)] exhibits an enhancement above a critical twist angle, at which there is a significant increase both in the 2D-peak width and blueshift. The above 2D-peak features are consistent with the results reported in twisted bilayer graphene[35] and signatures observed in t(2+2)LG[23], and are closely related to the twist-angle-dependent interlayer interaction and Van Hove singularities in tFLG. These results can also be used to confirm the validity of the twist-angle value determined from the included angle between the cracked edge and the folding boundary as discussed above. We now examine the G peak. It was demonstrated that the position of the G peak is highly sensitive to doping and strain in graphene.[36] We measured the Raman mapping of the G-peak position of the folding-twisted graphene samples [see Fig. 2(e) for the 11° t(2+2)LG as an example and Fig. S4 for more data]. No obvious shift in the G-peak position was observed by comparing the folded and unfolded regions of the samples. This indicates that the micro-tip-origami-prepared tFLG samples do not have significant further doping or strain. It is worth noting that the 2D/G peak intensity ratio also shows a twist-angle and layer-number dependence (see details in Fig. S5).[37]

In addition to the normal G and 2D peaks, new Raman modes marked by R and R' peaks emerge in our tFLG [Fig. 2(c)]. It was demonstrated that the twist induces superlattice wave vectors, resulting in the R and R' new Raman processes (transversal and longitudinal optical phonon branches) in twisted graphene, with their frequencies depending on the twist angle. Therefore, these R/R' peaks are strictly defined by the rotational angles, which have been widely reported in twisted bilayers.[21,38-40] The rich types of samples created here thus allow us to access the Raman characteristics of the R and R' modes in tFLG. The representative Raman spectra that contain R/R' peaks for t(2+2)LG are shown in Fig. 2(c). The R/R' peaks can be observed in various folded-

t(2+2)LG samples. From the R-peak intensity mapping [Fig. 2(e) and see Fig. S4 for more data], we can find that this peak only appears in the folded region, confirming the existence of a finite twist angle. Note that we only found partial tFLG samples showing obvious R/R' peaks. This is because the R and R' peaks are nondispersive under different laser excitations, and their intensities correlate with the excitation energy.[41] Figure 2(h) summarizes the R/R' modes frequencies as function of twist angle in our t(2+2)LG. The twist-angle dependence of R/R' modes can be clearly observed and shows good agreement with the theoretical predictions in twisted bilayer graphene.[22] This result provides a valuable reference for understanding superlattice-activated Raman processes.

The different stacking orders can be induced not only by interlayer twisting but also by interlayer slipping in graphene layers. For example, an interlayer slipping in the common ABA-stacked multilayer graphene could result in a metastable graphene structure, i.e., the ABC-stacked order. The ABC-stacked FLG has recently garnered considerable attention for its exotic electronic properties arising from the intrinsic flat band structures.[42-49] Using our polymer micro-tip-origami technique, we can create ABC-stacking-related tFLG. Figure 3(a),(b) shows a folding-generated t(3+3)LG ($\theta \sim$ 25°) from a ABC and ABA coexisting trilayer graphene—a ABC region folded on a ABC-ABA domain wall region. This generates various stacking configurations in one single sample, providing an unprecedented platform that could enable a direct comparative study between differently stacked graphene layers. There are four types of stacking configurations in this sample, i.e., ABC trilayer, ABA trilayer, twisted ABC/ABC double-trilayer and twisted ABC/ABA double-trilayer. Figure 3(c),(d) shows the representative Raman spectra of these four stacking regions, with the corresponding Raman maps of characteristic peaks displayed in Fig. 3(e)-(h). Obviously, there is a rotation-induced R peak for both the twisted ABC/ABC and ABC/ABA regions. The R peaks in these regions are located at the same position ($\sim$1380 cm$^{-1}$), suggesting that the ABC or ABA stacking sequence did not affect the twisting-activated Raman processes and confirming its superlattice origin.

Simultaneously, we investigated the 2D-peak line shape of the four stacking configurations. It was demonstrated that the ABC order exhibits a larger width and a higher asymmetry of the 2D peak compared to the ABA order.[50-53] This spectroscopic feature has been widely used to distinguish ABC stacking from ABA, as we applied here. As shown in Fig. 3(h), the 2D-peak FWHM of the ABC region (~67 cm$^{-1}$) is obviously larger than the ABA region (~63 cm$^{-1}$). Interestingly, the 2D-peak width of the twisted ABC/ABA (~68 cm$^{-1}$) is close to that of the ABC trilayer. This may be attributed to the relatively large angle of ~25° of this sample and the weak coupling between the upper and lower graphene layers.[23] Surprisingly, the 2D-peak FWHM of the twisted ABC/ABA (~64 cm$^{-1}$) is smaller than that of the ABC region. We speculate that this phenomenon may be due to the suppression effect from the underlying ABA graphene, as the Raman spectroscopy has a finite detection depth. This result can be supported by comparison of the 2D-peak shapes of these regions. As shown in the Raman spectra of Fig. 3(c),(d), the 2D-peak shape of ABC/ABC stacking is analogous to that of trilayer ABC stacking, while the 2D-peak shape of ABC/ABA stacking is more like ABA stacking.

We further examined the Raman G peak of the four stacking regions. A red shift of the G-peak position can be observed between the ABC and ABA trilayers (~2 cm$^{-1}$), as well as between the ABC/ABC and ABC/ABA twisted regions (~1 cm$^{-1}$). This observation aligns with previous Raman spectroscopy studies in ABC- and ABA-trilayer graphene, and has been attributed to subtle variations in their phonon band structures.[52] Interestingly, while the G-peak positions of ABC/ABA and ABA regions are nearly the same, there exists a clear blue shift (~1 cm$^{-1}$) in the twisted ABC/ABC double-trilayer compared to the ABC trilayer. This implies that the phonon band structure of ABC-stacked graphene layers is more sensitive to twisting.

We now study the structural stability of the folded-tFLG against thermal and mechanical perturbations. Figure 4 shows the thermal stability measurements of the folded-t(2+2)LG obtained by heating the samples at different temperatures for various durations. The folded-tFLG structure exhibits high thermal stability against annealing. Figure 4(a),(b) shows the G-Peak integrated intensity maps for two folded-t(2+2)LG samples annealed at 240 °C up to 36 h and for 12 h up to 500 °C, respectively. The twist angle and the whole structures of both samples exhibit no detectable changes from the analysis of the Raman and geometric characteristics [Fig. 4(c),(d) and see Figs. S6-9 for more data including AFM measurements and for more discussion about the edges]. Figure 4(e)-(g) summarizes the results of whether the folded-t(2+2)LG has changed after different annealing processes. Among 42 folded-t(2+2)LG samples we measured, 5 samples were changed in angle or area. We found that the folded-tFLG is more stable at annealing temperatures <300 °C—only 2 out of 30 samples underwent changes. Note that the two changed samples exhibit relatively small twist angles and sizes, which may exacerbate their instability. Besides, for all the 5 changed samples, their transformations occurred only once and all occurred during the first annealing [Fig. 4(h)]. We thus speculate that external factors, such as the presence of unreleased strain and impurity removal (see Fig. S10), may also contribute to the metastability of the folded-tFLG.

We have further checked the structural stability against mechanical perturbations through the transfer processes of the folded-tFLG. Figure 5 shows two folded-tFLG samples—a (2+2) and a (10+10)—before and after being transferred onto the hBN substrates. Likewise, no detectable changes in twist angle and size have been observed, indicating the high structural stability of the folded-tFLG against mechanical transfers. We also performed transfer operations for other types of folded-tFLG and observed similar excellent transferability of the samples (see Fig. S11). These results suggest the excellent stability and accessibility of the folded-tFLG for further device fabrications.

We finally discuss the possible mechanism for the structural robustness of the origami-created tFLG. It is well known that the twisted structure represents a metastable state in van der Waals layered materials, particularly at small twist angles.[54] However, the folded-tFLG exhibits several structural differences from the normal twisted graphene, which may account for its well stability. Firstly, there is a curved folding boundary connecting the two rotational layers in the folded-tFLG [see Fig. 2(a)]. This curved folding boundary, which possesses certain curvature energy,[55,56] may exhibit robust connectivity and effectively prevent the structural change of the folded-tFLG. In addition, most of the folded-tFLG exhibits tearing. The tearing edge of the folded layers will contain many dangling bonds, which may form additional connections with the underlying twisted layers through the formation of new C-C bonds.[57] This could further lock the adjacent folded graphene layers, enhancing the robustness of the formed tFLG structure.

In summary, we have presented a simple and efficient method to fabricate tFLG based on a polymer micro-tip-manipulated origami technique. We prepared various types of tFLG with different twist angles (0°-30°), including twisted bilayers (1+1), twisted double-bilayers (2+2), twisted double-trilayers (3+3), twisted thicker layers, and even twisted ABC multilayers. These origami-fabricated tFLG exhibit high structural stability against annealing and transferring processes, establishing folded-tFLG as a promising platform for high-stability device applications in twistronics. In addition, the rich variety of twisted structures also allows us to unveil the twist-angle and stacking-order dependent Raman characteristics of the tFLG, providing valuable insights into superlattice-induced phonon spectroscopy in graphene. Our developed polymer micro-tip manipulation technique may also inspire further applications in other van der Waals layered materials, promoting controllable manipulations of 2D materials into multiple configurations and dimensionalities.

**Acknowledgements**

This work was supported by the National Natural Science Foundation of China (Grant Nos. 12174095, 12474166, 12174096, 62101185, 12204164 and 51972106), and the Natural Science Foundation of Hunan Province, China (Grant No. 2025JJ20001). L.-J.Y. also acknowledges support from the Science and Technology Innovation Program of Hunan Province, China (Grant No. 2021RC3037). The authors acknowledge the financial support from the Fundamental Research Funds for the Central Universities of China.

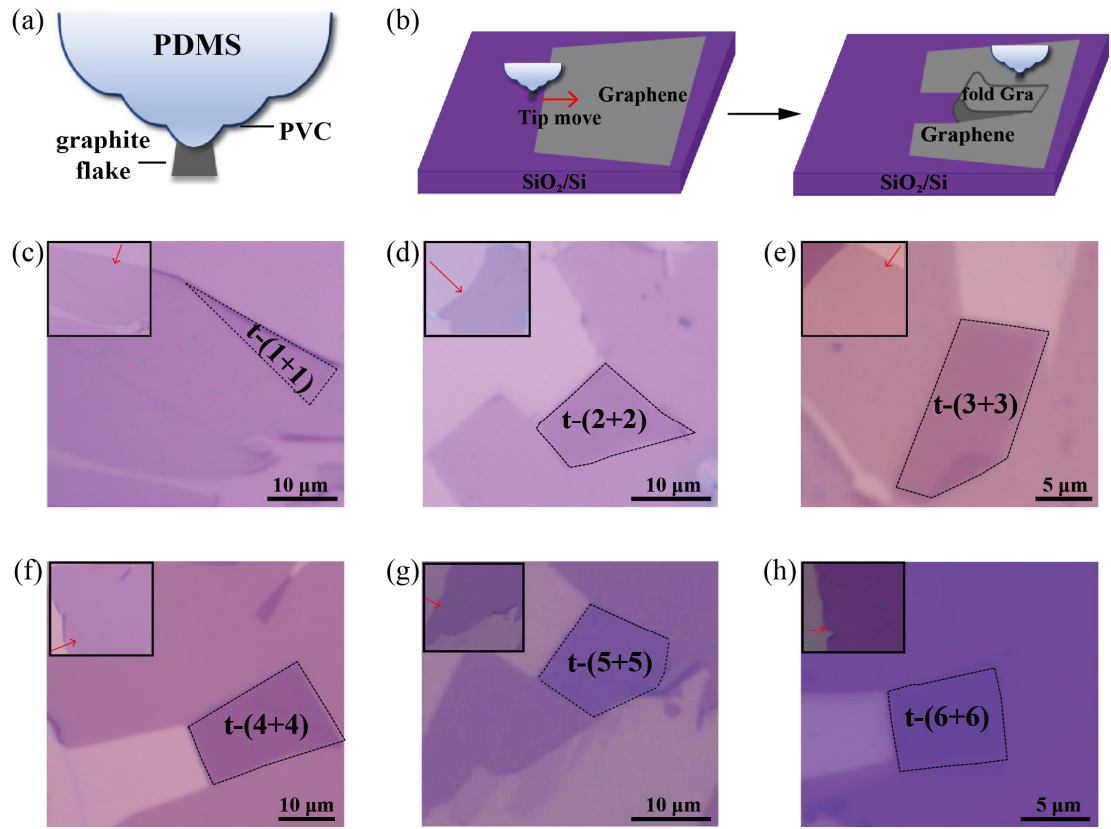

**FIG. 1**. (a) Schematic of the self-prepared polymer micro-tip. (b) Schematic illustration of the polymer micro-tip manipulated folding of graphene. (c-h) Optical images of the micro-tip origami created t(1+1)LG (14°; c), t(2+2)LG (11°; d), t(3+3)LG (25°; e), t(4+4)LG (20°; f), t(5+5)LG (12°; g), and t(6+6)LG (21°; h) samples. Insets show the optical images of the corresponding graphene samples before folding. The red arrows denote the direction of tip movement.

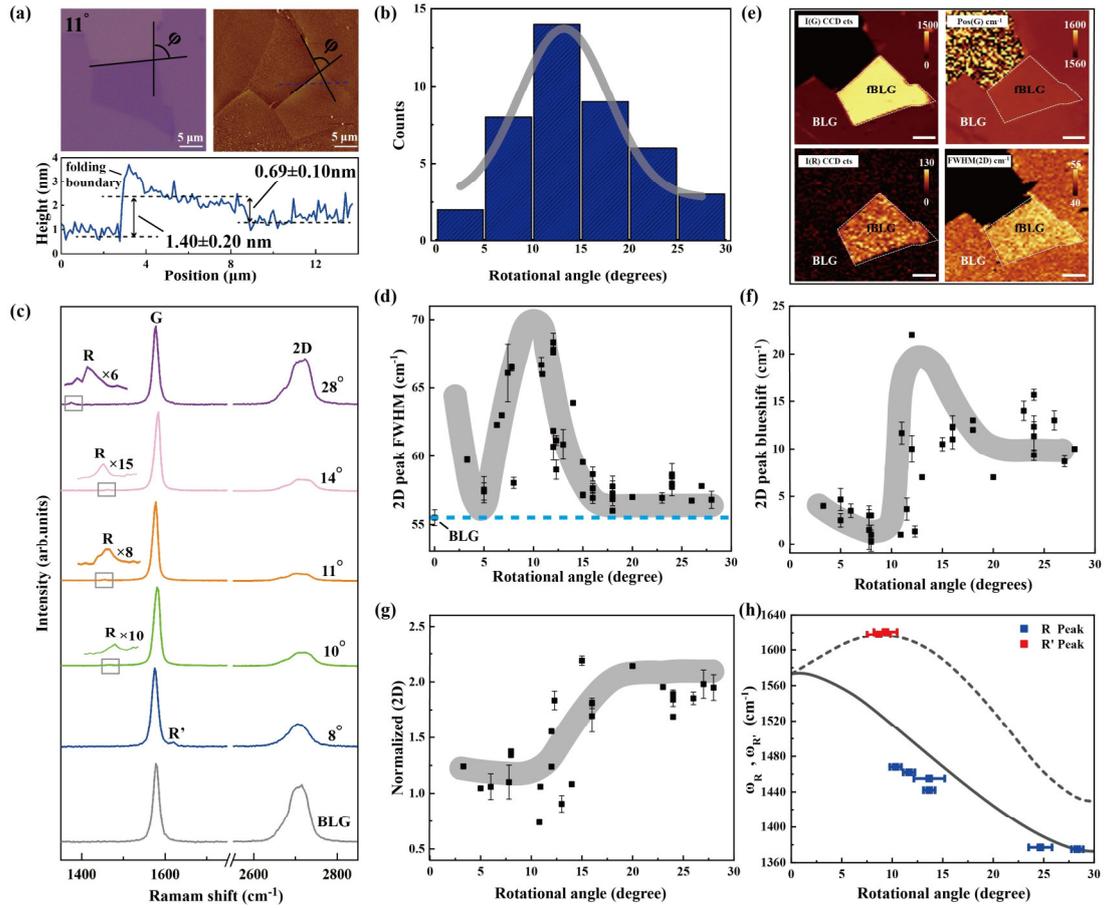

**FIG. 2**. (a) Optical and AFM images (upper panels) of a 11° t(2+2)LG sample, showing the extraction of the twist angle. The lower panel shows the height-profile across the folded region along the dashed line in the AFM image. (b) Statistical diagram of the twist angle for the micro-tip-origami created t(2+2)LG samples. The gray line is the gaussian fitting curve of the data. (c) Representative Raman spectra of the t(2+2)LG samples with different twist angles. (d) The FWHM of the Raman 2D peak as a function of rotational angle for the t(2+2)LG samples. (e) Raman maps of the G peak integrated intensity and position, R peak integrated intensity and FWHM of 2D peak for the 11° t(2+2)LG sample shown in (a). The scale bars denote 5 μm. The excitation energy is $E_{laser}$ = 2.54 eV (488 nm). (f,g) The blueshift (f) and normalized integral intensity (g) of the Raman 2D peak, with respect to the value of bilayer graphene, as a function of rotational angle for the t(2+2)LG samples. The gray bold line in (d), (f), (g) is a guide to the eyes. (h) R and R' peak frequency as a function of twist angle obtained from the t(2+2)LG samples. The dotted and solid curves are the theoretical results of R and R' peak frequencies of twisted bilayers taken from ref. 22.

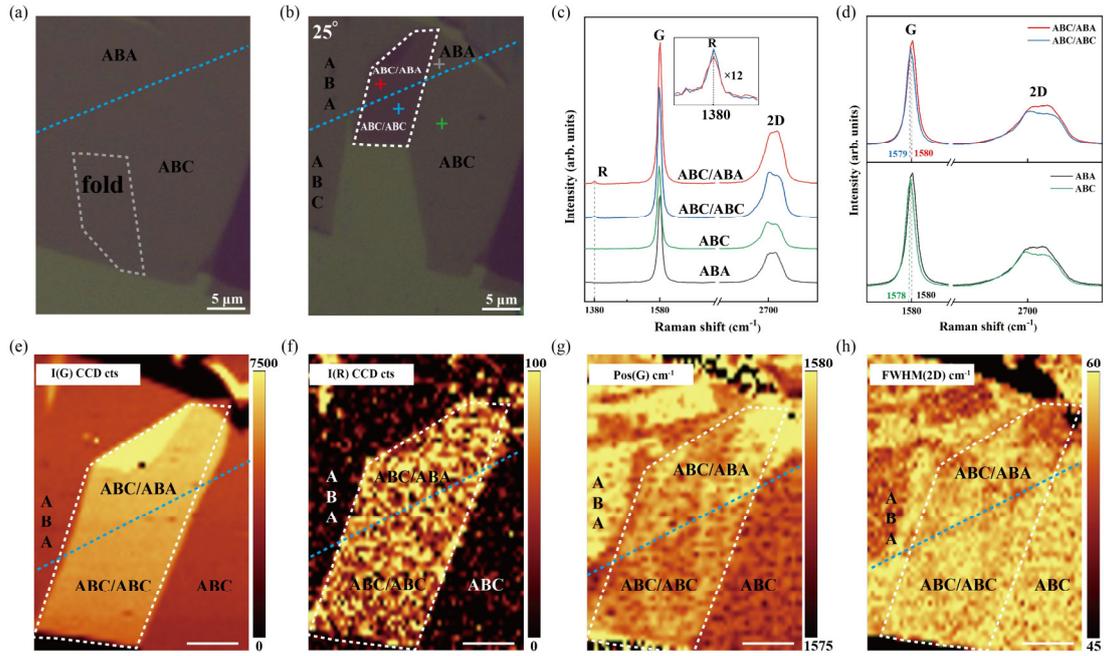

**FIG. 3**. (a,b) Optical images of a 25° t(3+3)LG sample before and after micro-tip manipulated folding. (c,d) The Raman spectra of four different stacking regions recorded at the positions marked by colored crosses in (b). (e-h) Raman maps of the G peak integrated intensity (e) and position (g), the R peak integrated intensity (f) and the 2D peak FWHM of the 25° t(3+3)LG sample shown in (b). The folding-twisted region can be clearly distinguished from the Raman intensity maps of G (e) and R peaks (f). The scale bars are 3 μm. The excitation energy is $E_{laser}$ = 2.54 eV (488 nm).

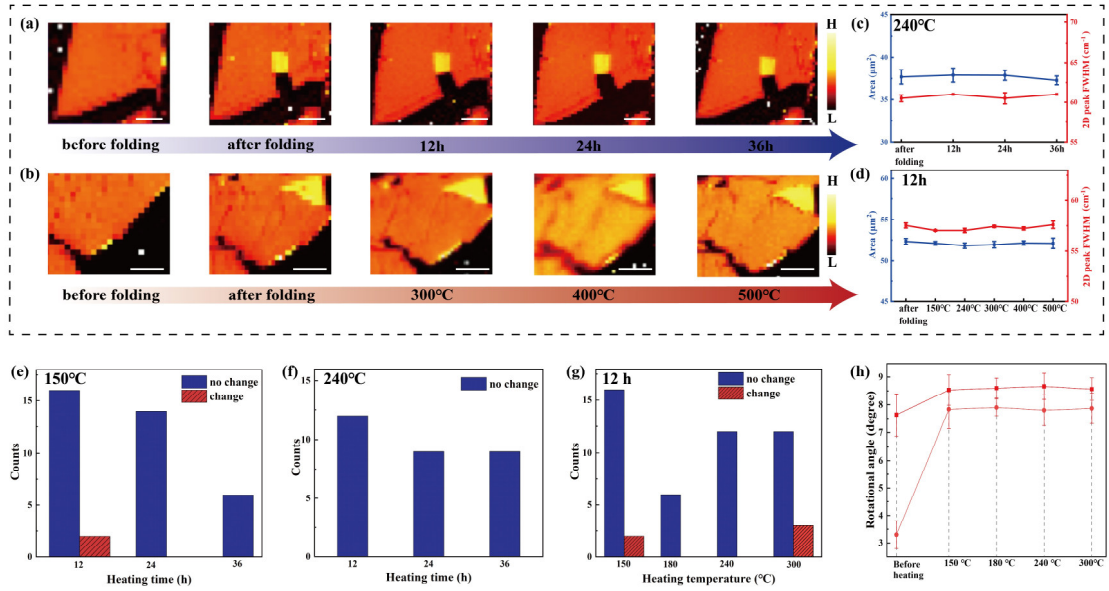

**FIG. 4**. (a) Raman images of a t(2+2)LG sample heating at 240 °C for 12 h, 24 h, 36 h. (b) Raman images of another t(2+2)LG sample heating at 300 °C, 400 °C, 500 °C for 12 h. Scale bars in (a) and (b) are 10 μm. (c,d) Extracted area and 2D-peak FWHM during different heating periods for the t(2+2)LG samples shown in (a) and (b), respectively. The excitation energy is $E_{laser}$ = 2.33 eV (532 nm). (e,f) Statistical diagram of twist-angle change in t(2+2)LG samples heating at 150 °C (e) and 240 °C (f) for different time. (g) Statistical diagram of twist-angle change in t(2+2)LG samples heating 12 h at different temperatures. (h) The angle evolution of the two changed t(2+2)LG samples as a function of heating temperature.

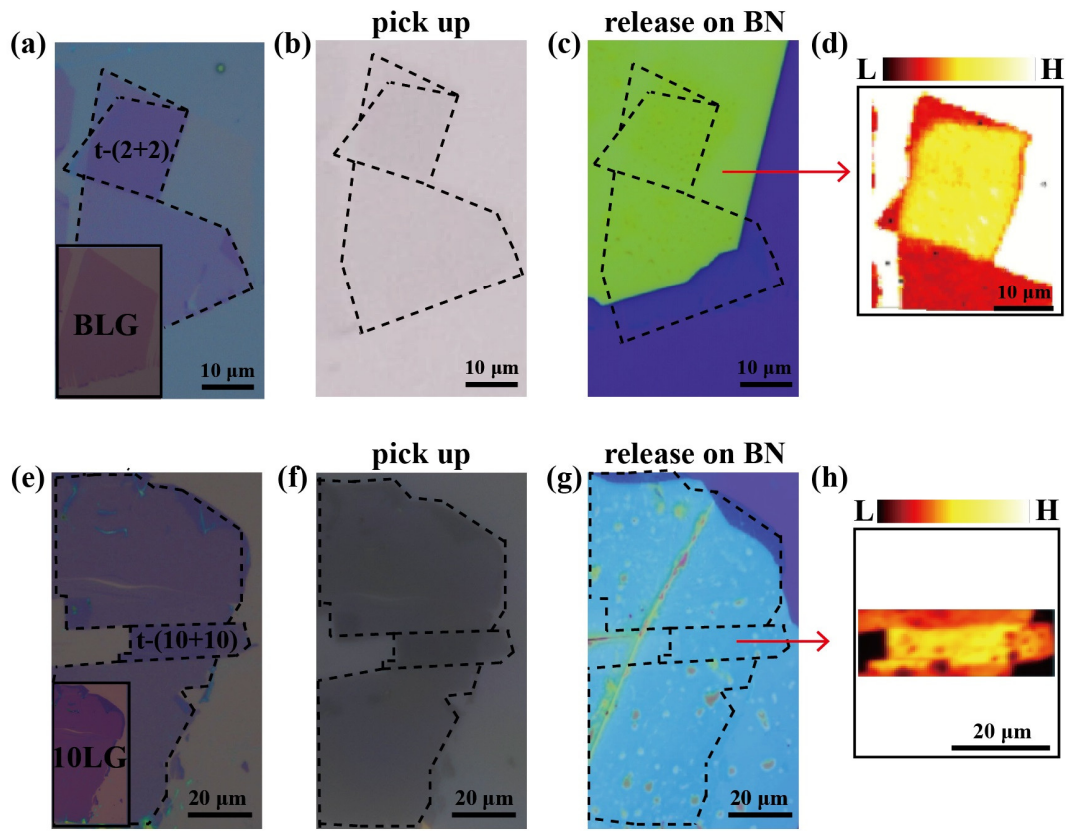

**FIG. 5**. (a-c) Optical images for the transfer process of a t(2+2)LG sample. (a) Before picking up on SiO$_2$; (b) After picking up by PVA; (c) After releasing on hBN. (d) Raman 2D-Peak integrated intensity map of the folded-t(2+2)LG region in (c). (e-g) Optical images for the transfer process of a t(10+10)LG sample. (e) Before picking up on SiO$_2$; (f) After picking up by PVA; (g) After releasing on hBN. (h) Raman G-Peak integrated intensity map of the folded-t(10+10)LG region in (g). Insets in (a) and (e) show the original graphene sheets before folding.